\documentclass[journal]{IEEEtran}
\usepackage{graphicx}
\PassOptionsToPackage{hyphens}{url}
\usepackage{hyperref}
\ifCLASSINFOpdf
\else
\fi
\begin{document}
\title{Blockchain-based Privacy Preservation for 5G-enabled Drone Communications}

\author{Yulei Wu,~\IEEEmembership{Senior Member,~IEEE,}
        Hong-Ning Dai,~\IEEEmembership{Senior Member,~IEEE,}
        Hao Wang,~\IEEEmembership{Member,~IEEE,
        Kim-Kwang Raymond Choo,~\IEEEmembership{Senior Member,~IEEE}}% <-this % stops a space
\thanks{Y. Wu is with the College of Engineering, Mathematics and Physical Sciences, University of Exeter, Exeter, EX4 4QF, U.K. e-mail: y.l.wu@exeter.ac.uk (Corresponding author)}% <-this % stops a space
\thanks{H.-N. Dai is with the Faculty of Information Technology, Macau University of Science and Technology, Macau. email: hndai@ieee.org}% <-this % stops a space
\thanks{H. Wang is with the Department of Computer Science, Norwegian University of Science and Technology, Gj\o vik, Norway. email: hawa@ntnu.no}% <-this % stops a space
\thanks{K.-K. R. Choo is with the Department of Information Systems and Cyber Security, University of Texas at San Antonio, San Antonio, TX 78249-0631, USA. email: raymond.choo@fulbrightmail.org}% <-this % stops a space
}


\markboth{}%
{}



\maketitle

\begin{abstract}
5G-enabled drones have potential applications in a variety of both military and civilian settings (e.g., monitoring and tracking of individuals in demonstrations and/or enforcing of social / physical distancing during pandemics such as COVID-19). Such applications generally involve the collection and dissemination of (massive) data from the drones to remote data centres for storage and analysis, for example via 5G networks. Consequently, there are security and privacy considerations underpinning 5G-enabled drone communications. We posit the potential of leveraging blockchain to facilitate privacy preservation, and therefore in this article we will review existing blockchain-based solutions after introducing the architecture for 5G-enabled drone communications and blockchain. We will also review existing legislation and data privacy regulations that need to be considered in the design of blockchain-based solutions, as well as identifying potential challenges and open issues which will hopefully inform future research agenda.
\end{abstract}

\begin{IEEEkeywords}
Drones, 5G, Blockchain, Privacy Preservation, Legislation and Data Privacy Regulations.
\end{IEEEkeywords}

% \fi
\IEEEpeerreviewmaketitle

\section{Introduction}
Drones (also referred to as unmanned aerial vehicles in the literature) are gaining popularity in a wide spectrum of tasks, ranging from military settings (e.g., reconnaissance and observation) to civilian scenarios (e.g., supporting search and rescue operations, monitoring weather and traffic flows, delivering goods, aerial photography and civilian monitoring and surveillance to enforce stay-at-home or social / physical distancing orders, for example during pandemics such as COVID-19), and so on~\cite{Tezza-2019}. Generally in these tasks, massive volume of data are collected and transferred to remote data centres for storage and analysis; thus, resulting in potential security and privacy challenges for both individuals and businesses. Individual privacy violations, for example, include being the subject of a targeted, but unauthorised (i.e., non-court approved) surveillance or some general-purpose aerial photography activities. There are also potential national security implications associated with drone activities, for example using drones to take pictures and videos of key military / sensitive installations. 

The fifth generation (5G) mobile communication system has been deployed in many countries, such as Australia, US, UK and China, to support a wide range of applications with diversified requirements (e.g., ultra-high bandwidth, ultra-low latency and ultra-high reliability). A swarm of 5G-enabled drones, coordinating and collaborating with each other, can form a web of networking, computing and storage resources in the sky. These flying resources facilitate sensing, analysing, and transmitting of collected data, particularly for drones equipped with high definition cameras and 5G-enabled communication modules \cite{Yang-TMC-2019}. These drones can also work with other 5G components (e.g., edge computing servers) to enhance their computing and storage capabilities. There are, however, underlying privacy considerations associated with the data collection, handling, storage and analysis.

\begin{figure*}[ht]
	\includegraphics[width=\textwidth]{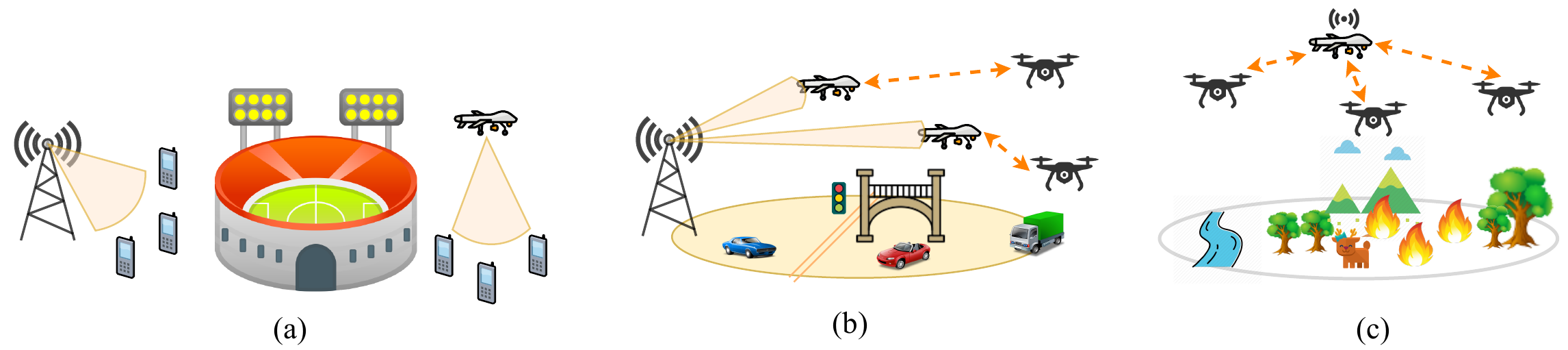}
	\caption{Three typical scenarios in 5G-enabled drone communications: (a) Using 5G base station drones to enhance 5G communication capacity, (b) Using 5G user drones to draw a map in a drone-to-drone (D2D) cooperative way, and (c) Using a combination of both 5G base station drones and 5G user drones to perform search and rescue in a wildfire.}
	\label{drone-scenario}
\end{figure*}

Privacy breaches can be examined from two perspectives. First, drones if controlled by a malicious user or successfully compromised and taken over by a malicious user, can be easily (ab)used as a surveillance device to track and/or monitor individuals. Efficient authentication of controlling drones is, therefore, an important factor to minimise privacy breaches. Second, the transparency of data handling also requires further study. For example, many countries have their own data protection regulations, such as the General Data Protection Regulation 2016/679 (GDPR) used by countries in the European Union and the European Economic Area. Under GDPR, personal private data need to be handled in the way agreed by the data owner. Ensuring the transparency of data handling is crucial in ensuring compliance with data protection regulation and minimising privacy breaches. There have been attempts to leverage blockchain to facilitate privacy preservation, as we will explain in this article.

In this paper, we will introduce the architecture for 5G-enabled drone communications, and briefly review the workings of blockchain and how it can facilitate privacy preservation (see Sections \ref{section:5G} and \ref{section:Blockchain for Privacy Preservation}, respectively). Then, in Section \ref{section:Review}, we will review blockchain-based solutions that can be adopted for 5G-enabled drone communications to minimise privacy breaches. In Section \ref{section:Privacy-related Legislation and Standards}, we will revisit existing legislation and data privacy regulations that need to be considered in the design of blockchain-based solutions. Finally, potential challenges and open issues associated with privacy preservation for 5G-enabled drone communications will be discussed.

\section{5G-enabled Drone Communications} \label{section:5G}
In 5G-enabled drone communications, drones typically act in two capacities, namely: 5G base stations/relays and 5G users \cite{Zeng-IEEE-Proceedings-2019}. Existing drones can be equipped with a lightweight base station or a relay. In other words, they become 5G base station/relay drones that facilitate terrestrial wireless communications. Such settings can be deployed in many different 5G application scenarios, such as at events (e.g., concerts) where spikes in traffic are expected only during certain times, and natural disasters (e.g., forest / bush fires) where there is no supporting infrastructure or the infrastructure is damaged. In addition, drones that act as 5G base stations/relays usually provide more reliable line-of-sight (LoS) connection links with their ground users compared with their terrestrial counterparts. 

Drones can also be the users of 5G systems, in order to leverage the features of 5G systems (e.g., ubiquitous coverage, low-latency and high-bandwidth). In such a setting, drones are controlled by ground stations to perform the allocated tasks. In addition, device-to-device communications in 5G systems allow drones to communicate with each other in an ad-hoc manner without the need of an (expensive) infrastructure. For example, drones can swiftly form a web of flying resources at the region in the immediate aftermath of a disaster, providing the necessary communication, storage and computing resources to facilitate activities such as rescue and search. 

In addition to the above two broad scenarios, drones can also play the role of both 5G base stations/relays and users in the same deployment, as demonstrated in Fig. \ref{drone-scenario}. Drones in the first two scenarios generally collect and/or produce massive volume of data, including sensitive  (e.g., videos or images relating to a key installation or suspects). Consequently, this raises the issues of who can legitimately access the data (and how to enforce such access control), and how to protect the privacy of the data.

\subsection{Potential Privacy Concerns}
As previously discussed, privacy is an ongoing concern~\cite{Lin2018}. In deployments where drones are the 5G base stations/relays (e.g., in public safety or crowd control situations, such as demonstrations or riots), the collected (sensitive) data can be targeted by (politically or issue-motivated) attackers seeking to exfiltrate the data. When drones are 5G users (e.g., in civilian monitoring and surveillance), attackers may seek to compromise and take over control of the drones and used them for nefarious purposes (e.g., as weapons to carry out attacks against the crowd \cite{DBLP:journals/fgcs/ParraRC20}). Data acquired by other drones in the vicinity could be eavesdropped by these compromised drones, for example by abusing device-to-device communications.

\section{Blockchain for Privacy Preservation} \label{section:Blockchain for Privacy Preservation}

\begin{figure}[ht]
    \centering
	\includegraphics[scale=0.6]{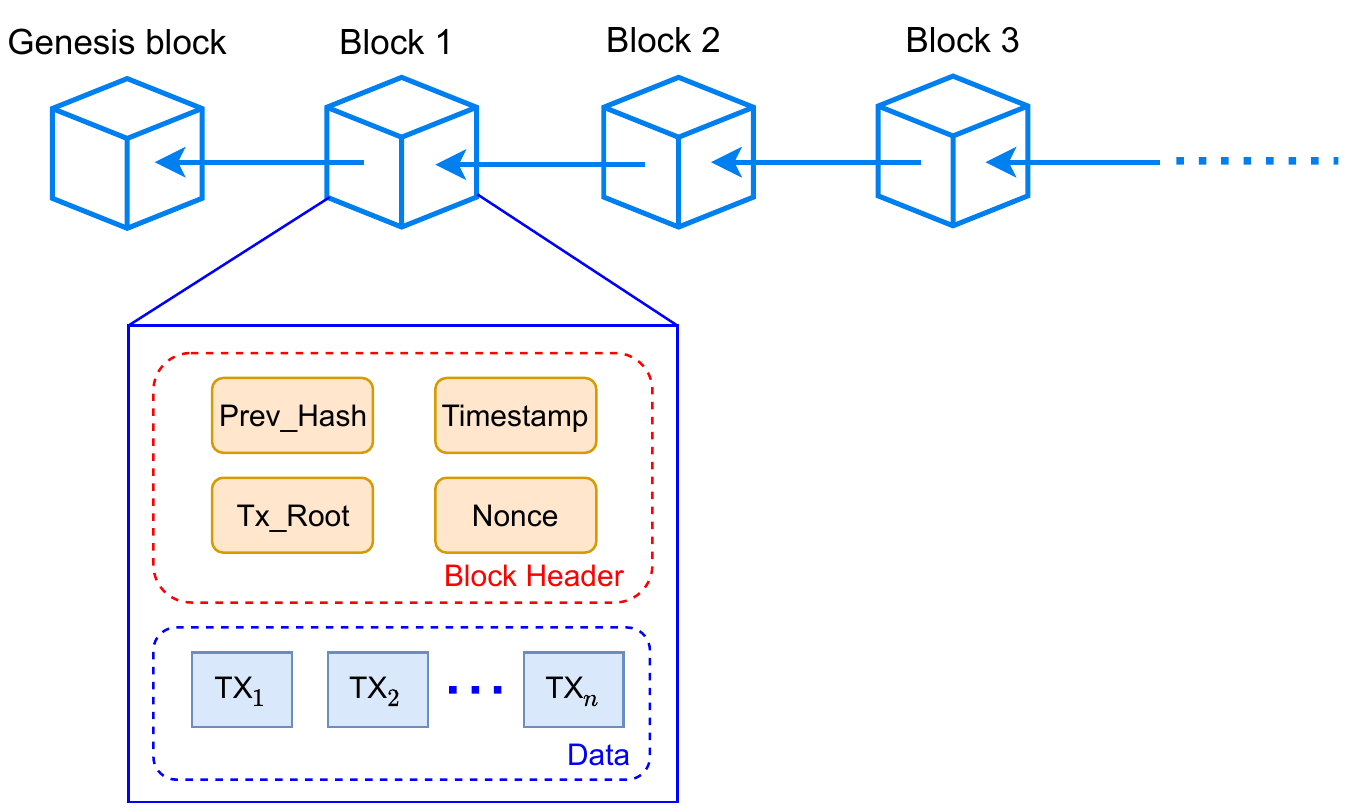}
	\caption{A blockchain typical structure.}
	\label{blockchain-structure}
\end{figure}

Blockchain is a decentralised distributed ledger database system, which contains cryptographically generated data blocks, where each block comprises a series of transactions approved by the majority of the participants in the system~\cite{Cheng-TCSS-2019}. Blocks are chained together (hence, the name ``blockchain'') in a linear fashion and in a chronological order. Each block possesses a hash of its previous block (recorded in the block header), which is used as the unique identification of the block. The hash value in a block is the hash value of its parent block, and a block in the blockchain can therefore be found through the hash value of its parent block. A chain with the linked list data structure, as shown in Fig.~\ref{blockchain-structure}, is formed by the hash value sequence of each block linked from the last generated one to the genesis one.

The following characteristics of blockchain can be utilized to facilitate privacy preservation:
\begin{itemize}
\item \textit{Transparency}. Each participant in the blockchain system can hold a copy of the blockchain; thus, allowing each participant to verify whether a transaction is initiated by a legitimate user.
\item \textit{Temper-proof}. Each block is added to the blockchain through the confirmation by a consensus algorithm, which undergoes a verification process of blocks where all participants can take part in. The blockchain system, therefore, maintains a tamper-proof ledger shared by the participants without relying on a trusted third party. 
\item \textit{Security}. Blockchain utilizes asymmetric cryptographic building blocks to encrypt data, whose security generally relies on the underpinning consensus algorithm (empowered by the majority of the participants).
%\item More features can be added if discussed in the next section.
\end{itemize}

Smart contracts \cite{8847638} are a key component of blockchain, enabling self-execution of a program when certain terms are met. Hence, they can be used to facilitate automated privacy preservation.

\section{Review of Blockchain-based Privacy Preservation Solutions} \label{section:Review}

We now present blockchain-based solutions for privacy preservation in 5G-enabled drone communications, in terms of ID management, data privacy protection, trajectory protection and the consensus of drone networks.

\subsection{Blockchain-based Identification Management of Drones}

\begin{figure*}[t]
\centering
	\includegraphics[scale=0.6]{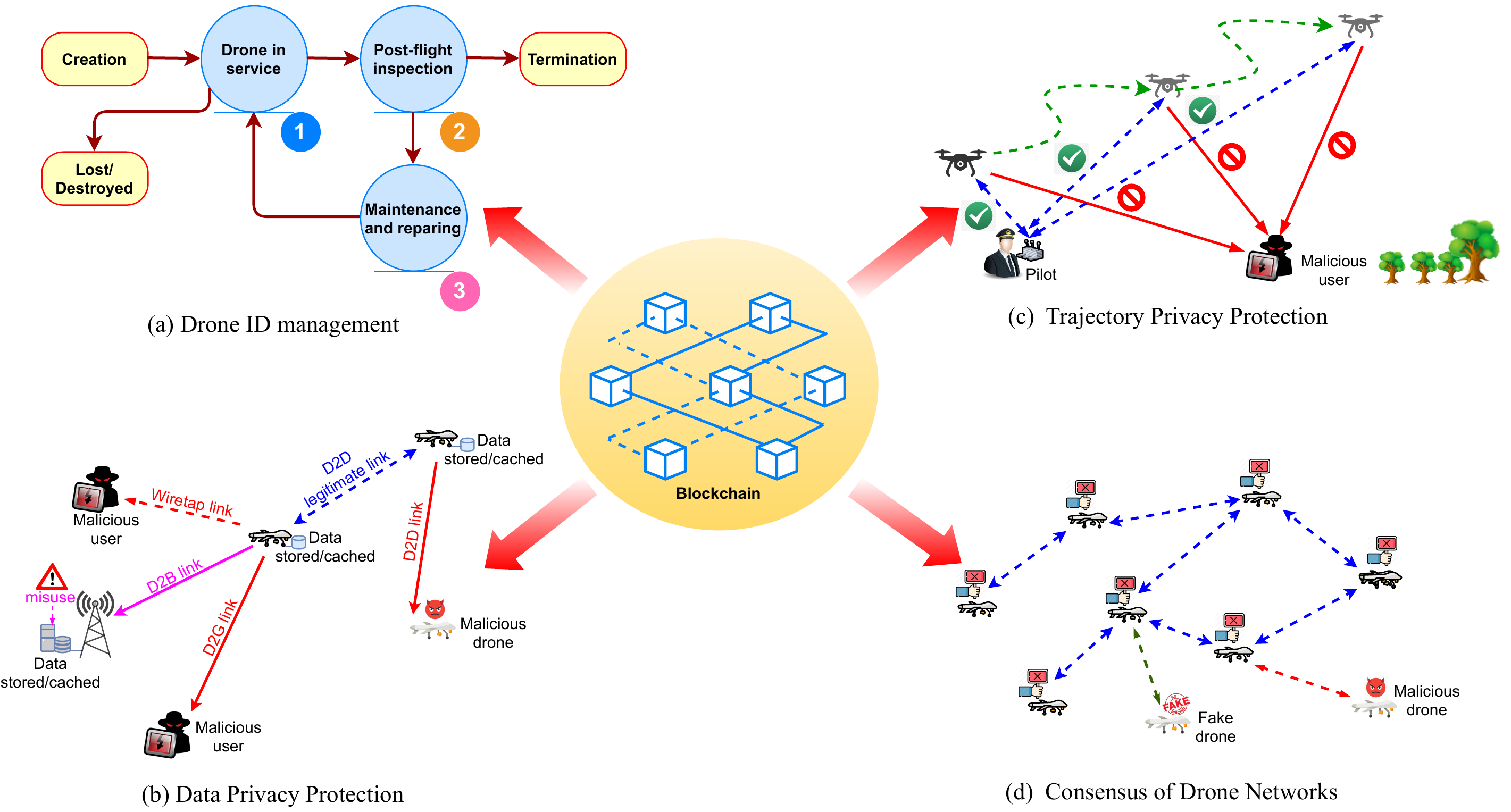}
	\caption{Blockchain-based Privacy Preservation for Drone Communications}
	\label{blockchain-drone}
\end{figure*}

Achieving effective and efficient identification (ID) management of drones is crucial as they are becoming increasingly commonplace. The importance of ID management is also reinforced by the guideline entitled ``Remote ID Notice of Proposed Rule Making (NPRM) for drone ID management''\footnote{\url{https://www.federalregister.gov/documents/2019/12/31/2019-28100/remote-identification-of-unmanned-aircraft-systems}} introduced by U.S. Federal Aviation Administration (FAA). 

However, centralised ID management can incur significant administrative costs (e.g., due to bureaucracy) and have other limitations such as single point of failure / attack. Hence, there have been interests to explore the utility of blockchain in designing decentralised ID management systems, in order to simplify the ID management process and lower the administration costs. In a robust blockchain-based system, for example, a drone can register and/or revoke any expired ID in the decentralised blockchain-based ID management systems. In addition, blockchain-based systems are tamper-proof, achieve non-repudiation, and minimise the single point of failure / attack risk. In addition, blockchains can also ensure the anonymity of drones, since drones only use the generated addresses to interact with each other in the system.

The blockchain-based ID management of drones can cover the entire life cycle of a drone, which consists of six main stages (including the three marked active stages) as shown in Fig.~\ref{blockchain-drone}(a). In particular, a drone can register its ID in the decentralised blockchain-based ID management system since the date of production (i.e., \emph{Creation}). During this process, the trust can be ensured via decentralised consensus of blockchain-based systems. A post-flight inspection needs to be performed after every flight (i.e., \emph{Drone in service}), as well as some maintenance or repairing tasks. Due to fatigue or attrition, a drone will terminate its service after a number of flights (i.e., \emph{Termination}). The termination status will be updated in the blockchain so that the drone ID can be either revoked or removed. For example, a lost or damaged drone also needs to be reported and recorded in the blockchain-based ID management system. In summary, during the entire life cycle of a drone, the status changes of a drone can be traceable via the blockchain-based ID management system. 

As discussed earlier, smart contracts can automate the drone ID management process. For example, conditions or terms of drone services can be written into contractual clauses in computer programs. When some conditions reach (e.g., a drone is revoked), the actions (corresponding to program statements) will be automatically triggered and executed (e.g., the drone becoming inactive). In this way, the drone ID management process can be simplified and the corresponding administrative cost can be reduced.

\subsection{Blockchain-based Data Privacy Protection of Drones}

As mentioned above, data can be collected by or (temporarily) stored in the drones (e.g., when drones act as 5G base stations, relays, or users). Such data can be of interest to attackers, who may seek to exfiltrate data stored or cached at these drones using a malicious / compromised drone, either via drone-to-drone (D2D) link or by drone-to-ground (D2G) link as shown in Fig.~\ref{blockchain-drone}(b). In addition, data-in-transit (from drone to base station) via the drone-to-base-station (D2B) link can potentially be intercepted or wiretapped by a malicious user. There is also the risk that the base station may be targeted. This reinforces the importance of both security and privacy preservation of user data in drone communications. 

Blockchain can play different roles in ensuring data privacy of drone communications. First, blockchain-based authentication mechanisms can verify whether an access request initiated from a drone is authorised. We remark that the authentication mechanisms should be fully integrated with blockchain-based ID management systems, with appropriate access control settings. To minimise the risk of data being misused at either the drones or the base stations, blockchain-based cryptographic schemes can be utilized~\cite{BCao:2019}, such as asymmetric encryption algorithms and homomorphic obfuscations. Third, blockchain-based encryption schemes across the entire network stack (e.g., physical, link and network layers) can also be deployed to mitigate threats at D2D, D2G and D2B links~\cite{Wu-IEEEWirelessCom-2019}. Furthermore, the adoption of blockchain can preserve the privacy of cache content and ensure trust among multiple parties as explained in~\cite{YQian:NetMag20}. 

In addition to data privacy protection, we also need to ensure efficient management of drone data (e.g., both blockchain data and user data). In particular, public blockchain systems are known to have massive blockchain data volume, e.g., Bitcoin contains more than 240GB data as of 2019. It is impractical to store the entire blockchain data at drones, even for higher-end drones with a larger storage capacities. Therefore, drones may only store user data, such as data from other Internet-of-Things (IoT) devices, and potentially partial blockchain data (e.g., hash values of blockchain transactions for verification purpose). Ground or base stations may store the entire blockchain data. Other than data storage challenges, there are also latency and bandwidth challenges associated with the massive data size. Hence, it becomes an essential to preprocess the data at drones since user data, especially IoT data, may contain duplicates, errors and noises. The miniaturisation of high-performance computing facilities and the rapid development of embedding devices can help us meet this emerging demand. In other words, drones may serve as edge computing nodes to complement remote clouds and collaborate with other edge computing nodes deployed at base stations. 

%Public addresses (psonydonytiy), privacy encryption, public description. Digital signature to verify the truth. zero knowledge proof.

\subsection{Blockchain-based Trajectory Privacy Protection}

The trajectory information of drones is crucial to facilitate and enforce control, route planning and navigation during adverse weather conditions or natural disasters. However, the trajectory information of drones is vulnerable to malicious attacks, as shown in Fig.~\ref{blockchain-drone}(c). In addition, centralised trajectory information management is susceptible to single point of failure / attacks, denial-of-service (DoS) attacks, and privacy breaches. For example, drones can be tracked, intercepted and even hijacked once the trajectories of drones are exposed to malicious users. Moreover, the behaviours of drone users can be tracked and inferred by analysing the trajectories of drones. Such data can also be used to facilitate other nefarious activities, such as stalking.

Similar to the earlier discussion, blockchain-based authentication and access control mechanisms can used to authorise access permissions of users to the drone trajectory data. Again, the integration of blockchain-based trajectory management with blockchain-based ID management and blockchain-based data management is crucial. Second, the decentralisation of blockchain-based drone trajectory management systems can also mitigate the risks of single point of failure / attacks associated with centralised systems. Moreover, incorporating other trajectory privacy-preservation schemes (e.g., $k$-anonymity scheme) can better improve the trajectory privacy protection of drones.

% Countermeasures:
% Blockchain-based solutions: 
% 1) Authentication and access control mechanisms (determine who can get what kind of trajectory information), also based on blockchain-based ID management and data management. 
% 2) Integration with other privacy-preservation schemes.

% important since ...
% Trajectory privacy concerns:
% Drone trajectory can be used to route planning, traffic control etc. The trajectory privacy of drones can also lead to 
% 1) drones are intercepted, controlled.
% 2) user behaviour can be deduced
\subsection{Blockchain-based Consensus of Drone Networks}

In some real-world scenarios (e.g., monitoring crowd movements in a demonstration), multiple drones may collaborate together to complete a complex task. In this case, one task is divided into a number of sub-tasks, each of which is completed by a drone. During this process, it is crucial to ensure a reliable drone network that can coordinate between multiple drones. However, it can be extremely challenging to maintain a dependable network due to the dynamic topology of drone networks, unreliable wireless communications between drones, and the potential of drones to be compromised or attacked~\cite{YANMAZ:Adhocnet-2018}. Take Fig.~\ref{blockchain-drone}(d) as an example, a malicious drone may be disguised as a legitimate drone to join the drone network so as to carry out malicious activities such as disrupting the drone network or wiretapping private data transmission between drones.

How can blockchain play a role here? First, blockchain-based ID management system can be used to identify non-member drones by analysing and tracking ID updating historical records. Second, the consensus mechanisms (e.g., Proof of Work and Practical Byzantine Fault-Tolerance~\cite{Dai:IoTJ2019}) in blockchain can help the majority of legitimate drones to be resistant to malicious attacks, such as Sybil attack. Third, the consensus of drone networks can significantly raise the cost of counterfeiting a fake drone or several fake drones; consequently, mitigating falsification risks. Moreover, the incentive mechanisms in blockchain systems can be leveraged to motivate drones into participating in the consensus of drone networks. Instead of employing digital currencies as the direct incentive, reputation credits might be more suitable in the drone network scenario. 

\begin{table}[ht]
\caption{Summary of blockchain-based solutions in drone communications}
\label{tab:summary}
\renewcommand{\arraystretch}{1.5}
\begin{tabular}{|p{0.3cm}|p{3.6cm}|p{3.6cm}|}
\hline
 No. &  \textbf{Drone Communications} & \textbf{Blockchain solutions}  \\ \hline\hline
 1 & Drone ID management    &  Transparency, temper-proof, traceability   \\  \hline
 2 & Data privacy protection  &   Asymmetric encryption algorithms, homomorphic obfuscations     \\ \hline
 3 & Trajectory privacy      &  Authentication and access control schemes    \\ \hline
 4 &  Consensus of drone networks  &  Fault-tolerance, traceability, anti-falsification \\ \hline
\end{tabular}
\end{table}

Blockchain is not, however, a panacea for drone security and privacy. For example, many consensus algorithms of existing blockchains have low efficiency and incur significant resource overheads. Such limitations may limit the adoption of blockchain in drone networks. Directed acyclic graph (DAG), sharding blockchain consensus, off-chain blockchain data are possible solutions to these challenges. In particular, DAG accepts the non-conflict side-chain so as to reduce the cost, while the sharding consensus only requires a subset of nodes (corresponding to drones) to participate in the consensus procedure and multiple subsets (or committees) can then reach  consensus. Consequently, consensus efficiency can be greatly improved. In addition, off-chain strategies allow  transactions to be conducted without the involvement of the main blockchain, and all these transactions can be eventually stored as a new block to the main blockchain. 

Table~\ref{tab:summary} summarises the blockchain-based solutions in 5G drone communications.

\section{Privacy-related Legislation and Standards} \label{section:Privacy-related Legislation and Standards}
Similar to other technologies, drone-related regulations generally lag behind research and development advances in drones. For example, St{\"o}cker et al.~\cite{stocker2017} presented a comprehensive review of the status of drone-related regulations as of 2017. The surveys of Fotouhi et al.~\cite{fotouhi2019} and Ullah et al.~\cite{ullah2020} also discussed recent developments in drone-related regulations and standardisation. In this section, we focus on the regulatory and standardisation advances relating to privacy preservation in drone communications.

Public privacy, safety, and data protection are the key focuses in the majority of legislative efforts ~\cite{stocker2017}, for example to protect individuals, environment, and objects from the various harms (e.g., physical safety and intrusion of private space) associated with drones. While regulations may differ between jurisdictions, existing regulations tend to have clear definitions of no-fly zones, the need to maintain a safe distance from human crowds and prohibit the flying of drones over human crowds, the need for training and certification of pilots who fly drones over a defined weight limit, flying below a maximum flying height, and liabilities in the event of an incident. 

Examples of recently released regulations include the EU Commission Delegated Regulations 2019/945 (EU2019/945)~\footnote{Council  of  European  Union,  “Commission  Delegated  Regulation  (EU)2019/945 on unmanned aircraft systems and on third-country operators of unmanned aircraft systems,” 2019, https://eur-lex.europa.eu/legal-content/EN/TXT/?uri=CELEX:32019R0945.} and EU2019/947~\footnote{Council  of  European  Union,  “Commission  Delegated  Regulation  (EU)  2019/947  on  the  rules and procedures for the operation of unmanned aircraft,” 2019, https://eur-lex.europa.eu/legal-content/EN/TXT/?uri=CELEX:32019R0947.}, which are scheduled to be fully enforced on July 1 of 2020. These regulations cover the design, manufacture, and operation of drones, and has implications to manufacturers, importers, and distributors. In addition, the new regulations clarify on the technical requirements of drones in different classes. 

Another recent popular topic in the drone industry is the so-called \textit{Beyond Visual Line of Sight} (BVLOS) flights, which can cover larger areas (including areas that are difficult or impossible for the pilots to keep an eye on). BVLOS flights can be deployed in adversarial and rough conditions, such as battlefields, inspection of key installations (e.g., oil and gas pipelines, power grids, and border control) and wild life, and search and rescue operations. However, there are also greater risky or ill-intentioned use in the operating of BVLOS flights, which may explain why they are generally not allowed in many countries. For example, the U.S. does not allow BVLOS flights, without a waiver from the relevant authority\footnote{It is reportedly very difficult to obtain such a waiver in the U.S.. For example, as of June 8, 2020, only 54 BVLOS (107.31) waivers have been issued, according to https://www.faa.gov/uas/commercial\_operators/part\_107\_waivers/}. However, regulations on BVLOS flights are evolving at a very fast pace. An amendment\footnote{https://www.easa.europa.eu/sites/default/files/dfu/NPA\%202020-07.pdf} to EU2019/947 w.r.t. BVLOS flights is currently in progress, at the time of writing. We expect that the BVLOS flights will become better regulated across different countries in the near future, and it is an important aspect to consider when we study privacy preservation issues for drones.    

Standards bodies have also been very proactive in drone-related activities. For example, the technical specification TS 22.125 of 3GPP\footnote{TS 22.125 Unmanned Aerial System (UAS) support in 3GPP, 2019 \url{https://www.3gpp.org/uas-uav}} ``identifies the requirements for operation of UAVs via the 3GPP system''. The 3GPP Release 16 includes ``requirements for meeting the business, security, and public safety needs for the remote identification and tracking of Unmanned Aerial System (UAS) linked to a 3GPP subscription''. In the 3GPP Release 17 (scheduled for delivery in 2021), it includes 5G enhancement for UAVs. 

As observed by St{\"o}cker et al.~\cite{stocker2017}, an increase in drone activities will also result in additional administrative processes, such as those relating to flight registration and approval. We believe that the decentralisation of blockchain systems is a viable approach to reducing administrative redtapes. 

ID of drones has been one of the main artefacts to ensure traceability and accountability. In the new EU rules, with the exception of class C0 (less than 250g), all classes must bear a unique physical serial number and more importantly a \textit{direct remote ID} ``allowing the upload of the operator registration number and in real time during the whole duration of the flight, the direct periodic broadcast from the UA (unmanned aircraft) using an \textit{open and documented transmission protocol} in a way that they can be received directly by existing mobile devices within the broadcasting range''. On the other hand, TS 22.125 of 3GPP further elaborates that ``The 3GPP system shall enable UAV to preserve the privacy of the owner of the UAV, UAV pilot, and the UAV operator in its broadcast of identity information''. From these recent developments in regulations, it is clear that auditability and anonymity features due to the use of blockchain can facilitate traceability and accountability of drones.

On data privacy protection, we can look at two aspects. First, the privacy of people, environment, and objects that may be intruded by drones, and second, the protection of legitimate data collected by drones and the communication privacy between the drone and the pilot. As mentioned earlier, the first aspect has been the focus of recent legislative changes. However, in practice it can be difficult to enforce. The sensory range of onboard sensors is constantly improving due to technology advancement. This compounds the challenge of tracking and identification, especially for smaller drones. The direct remote ID and the geo-awareness system required by EU2019/945 for drones in some classes are helpful in this aspect, so further exploration is necessary. The second aspect is also partially covered by EU2019/945, since drones in some classes are required to ``be equipped with a data link protected against unauthorised access to the command and control functions'' and TS 22.125 of 3GPP states that ``3GPP system shall support the capability to provide different levels of \textit{integrity and privacy protection} for the different connections between UAS and UTM (UAS Traffic Management) as well as the data being transferred via those connections''.

Table~\ref{tab:regulations} presents a brief summary of regulations by selected representative countries\footnote{These countries are selected because they are representative of the most advanced development in drone regulations from different continents. St{\"o}cker et al.~\cite{stocker2017} also studied these countries in their comparative analysis, with the exception of the new EU regulations EU2019/945 and EU2019/947.}, in terms of privacy preservation. Recall that the new EU regulations EU2019/945 and EU2019/947 will be fully enforced on July 1 of 2020. Therefore, existing national regulations are in the process of being harmonised with or superseded by the new EU rules. The communication privacy on drones are largely not mentioned by the national regulations listed in Table~\ref{tab:regulations}. In other words, the new EU regulations and the 3GPP standards are more advanced in this aspect.

\begin{table*}[t]
  \begin{center}
    \caption{Comparison of drone-related regulations in terms of privacy preservation}
    \label{tab:regulations}
    \begin{tabular}{c|c|c} 
      \hline
      \textbf{Country} & \textbf{Data Privacy on Drones} & \textbf{\shortstack{Communication Privacy on Drones}}\\
      \hline
      \shortstack{EU 2019/945\\ EU 2019/947} & Legally regulated & \shortstack{Drones in some classes required\\ to be equipped with secure data link}  \\ \hline 
      Australia         & \shortstack{Only advice to respect private privacy\\ Privacy Act only applies on large organisations\\ Authority plans to review privacy issues with recreational drones} & N/A  \\ \hline
      Canada            & Privacy Act applies to commercial and government drones & N/A \\ \hline
      China             & \shortstack{Not in national laws\\ but covered by some provincial laws (e.g., Sichuan)}  & N/A \\ \hline
      Colombia          & Not allowed to violate the rights of privacy & N/A \\ \hline
      France            & Operators obliged to respect privacy rights of individuals  & \\ \hline
      Germany           & Bundesdatenschutzgesetz (BDSG, federal data protection act) applies  & N/A  \\ \hline
      Italy             & Italian Data Protection Code, enacting GDPR, applies  & N/A  \\ \hline
      Japan             & \shortstack{Not linked to the Act on the Protection of Personal Information (APPI)\\ but authority plans to cover privacy in next phase in the roadmap}  & N/A \\ \hline
      Rwanda            & \shortstack{Operators oblighed to respect privacy rights of others\\ surveillance of people and property without their consent is prohibited}  & N/A \\ \hline
      The Netherlands   & Operators not allowed to violate other people’s privacy  & N/A \\ \hline
      United Kingdom    & The Data Protection Act (DPA) applies  & N/A \\ \hline
      United States     & Covered differently by State- or City-level laws  & N/A \\
      \hline
    \end{tabular}
  \end{center}
\end{table*}

\section{Research Challenges and Open Issues}
%\subsection{Privacy preservation for drone communications in 5G}
Despite the potential benefits of blockchain in drone communication privacy preservation, there remain a number of open challenges which will be discussed next.

\begin{itemize}

\item \emph{Resource constraints of drones:} Most existing drones are resource-limited, in terms of energy, size and weight considerations. Encryption and/or consensus algorithms are generally required for blockchain systems, yet drones are generally incapable of computing-intensive tasks due to computationally constraints and battery life. In addition, a swarm of UAVs can generate and/or collect gigabytes of data per second, including both audio and video. Whether the storage capacity of blockchain can accommodate such high volume of data is still debatable, and whether and how to incorporate other storage resources (e.g., edge servers) with the UAV system remains an open challenge. Apart from these, drones are energy constrained devices, and thus they need energy efficient solutions. However, miners (i.e., drones) consume a disproportionate amount of electricity when generating blocks; thus, existing drones may not be capable of supporting sufficient energy required for mining of blocks. In the future, the orchestration of various computing facilities such as remote clouds, nearby edge servers and drones, and other technologies such as network coding, becomes a necessity to implement blockchain-based drone communications.

%\item \emph{Storage constraints:} A swarm of UAVs can generate and/or collect gigabytes of data per second, including both audio and video. Whether the storage capacity of blockchain can accommodate such high volume of data is still questionable, and whether and how to incorporate the cloud/edge storage with the UAV system is still an open issue.

%\item \emph{Energy consumption issues:} Drones are energy constrained devices, and thus they need energy efficient solutions. However, miners (i.e., drones in case of UAV systems) consume a disproportionate amount of electricity when generating blocks. The energy consumption for mining blocks is still a challenging issue to UAV systems.

%\item \emph{The convergence with other technologies:} The solutions with blockchain itself may not be suitable for drone networks due to above-mentioned computing, storage and energy-consumption issues. The convergence of blockchain with other technologies, such as cloud computing, edge computing, fog computing and network coding is inevitable, to reach a balance between computing, storage and energy-consumption, but this is still an open issue that needs to be well addressed by researchers.

\item \emph{Full privacy preservation of drone data:} In blockchain-based solutions for drone networks, each drone requires to store a copy of the data blocks (i.e., distributed ledger). This risks the dissemination of sensitive information to all participating drones. Although blockchain can guarantee certain level of privacy preservation of drone data, activities of both users of drone communications and drones can be inferred (or extracted) via statistical analysis or using other machine learning tools. For example, user private data relayed through drones may be leaked to malicious users who may compromise the drones with the aim of exfiltrating data. How to fully ensure data privacy of drone communications is still an open research question. Limiting the information sharing between drones is one potential solution, although this may not be practical in some applications.

\item \emph{Scalability of blockchain-based drone networks:} Multiple drones can form a drone network for diverse tasks. As discussed earlier, the consensus of drone networks can help to mitigate the falsification of malicious drones and other security risks. However, it is challenging to achieve a scalable blockchain-based drone network due to the dynamics of drones (i.e., drones can join and leave at any time) as well as the scalability constraints of current blockchain systems (i.e., low throughput of transactions per second). For example, poor scalability may lead to the difficulty of forming a drone network and reaching a consensus when a new drone joins. Therefore,  scalability of blockchain-based drone networks is an important issue to explore in the future.

\item \emph{Remote identification:} As mentioned earlier, new regulations require drones to periodically broadcast their ID information that can be directly received by existing mobile devices within the broadcasting range. Such an activity needs to be conducted without violating the privacy of the owner, the pilot, and the operator. Designing an efficient solution for remote ID requires an in-depth understanding of the data transmission protocols and the various security and privacy risks (including emerging risks), and hence remains one of ongoing interest.

\item \emph{Regulation development and compliance enforcement:} Regulations and standards for drones are still evolving, and privacy preservation remains an prioritised agenda. Drone accidents may occur due to a range of reasons, such as technical malfunction, improper operations, unforeseen environmental events (e.g., sudden wind gusts), and hijacking. As more automation functionalities are being introduced into drones, clear definitions of liabilities and responsibilities for all participants involved across the entire life cycle of a drone will need to be explored. A closely related issue is how to enforce the compliance of regulations for drones. The collection and certification of digital evidence (enabled by blockchain) on drone accidents or privacy intrusions / breaches are also potential research topics.

\end{itemize}

\section{Conclusion}
This article discussed blockchain-based privacy preservation solutions for 5G-enabled drone communications, as well as related data privacy legislation and regulations that need to be considered in the design of these solutions. We also identified potential challenges and open issues to inform future research agenda that will allow the community to leverage blockchain to facilitate privacy preservation in drone communications.

\bibliographystyle{IEEEtran}
\bibliography{example_bib}

\begin{IEEEbiographynophoto}{Yulei Wu}
[SM’18] is a Senior Lecturer with the Department of Computer Science, College of Engineering, Mathematics and Physical Sciences, University of Exeter, United Kingdom. He received the B.Sc. degree (First Class Honours) in Computer Science and the Ph.D. degree in Computing and Mathematics from the University of Bradford, United Kingdom, in 2006 and 2010, respectively. His expertise is on networking and his main research interests include computer networks, networked systems, software defined networks and systems, network management, and network security and privacy. He is an Editor of IEEE Transactions on Network and Service Management, Computer Networks (Elsevier) and IEEE Access.
\end{IEEEbiographynophoto}

\begin{IEEEbiographynophoto}{Hong-Ning Dai}
[SM’16] is currently with Faculty of Information Technology at Macau University of Science and Technology as an associate professor. He obtained the Ph.D. degree in Computer Science and Engineering from Department of Computer Science and Engineering at the Chinese University of Hong Kong. His current research interests include Internet of Things and blockchain technology. He has served as an associate editor for IEEE Access, guest editors for IEEE Transactions on Industrial Informatics, IEEE Transactions on Emerging Topics in Computing. He is a senior member of the Institute of Electrical and Electronics Engineers (IEEE). 
\end{IEEEbiographynophoto}

\begin{IEEEbiographynophoto}{Hao Wang}
[M’07] is an Associate professor in the Department of Computer Science in Norwegian University of Science \& Technology, Norway. He received his Ph.D. degree and a B.Eng. degree, both in computer science and engineering, from South China University of Technology, China in 2006 and 2000, respectively. His research interests include big data analytics, industrial internet of things, high performance computing, and safety-critical systems. He is a member of IEEE and ACM. He is the Chair for Sub TC on Healthcare in IEEE IES Technical Committee on Industrial Informatics.
\end{IEEEbiographynophoto}

\begin{IEEEbiographynophoto}{Kim-Kwang Raymond Choo}
[SM’15] received the Ph.D. in Information Security in 2006 from Queensland University of Technology, Australia. He currently holds the Cloud Technology Endowed Professorship at The University of Texas at San Antonio. In 2015, he and his team won the Digital Forensics Research Challenge organized by Germany's University of Erlangen-Nuremberg. He is the recipient of the 2019 IEEE Technical Committee on Scalable Computing (TCSC) Award for Excellence in Scalable Computing (Middle Career Researcher). He is also an Australian Computer Society Fellow, an IEEE Senior Member, and Co-Chair of IEEE Multimedia Communications TC's Digital Rights Management for Multimedia Interest Group. 
\end{IEEEbiographynophoto}
% or


%\appendices

%\section{}

%\section*{Acknowledgment}


\ifCLASSOPTIONcaptionsoff
  \newpage
\fi

\end{document}